\documentclass[11pt,a4paper]{article}

\usepackage[T1]{fontenc}
\usepackage[margin=1in]{geometry}
\usepackage{amsmath}
\usepackage{amssymb}
\usepackage{graphicx}
\usepackage{xcolor}
\usepackage{listings}
\usepackage{fancyhdr}
\usepackage{booktabs}
\usepackage{array}
\usepackage{tabularx}
\usepackage{multirow}
\usepackage[hidelinks]{hyperref}
\usepackage[numbers,sort&compress]{natbib}
\usepackage{setspace}
\usepackage{float}
\usepackage{caption}
\usepackage{enumitem}
\usepackage{ragged2e}
\usepackage{etoolbox}
\usepackage{titlesec}

\titlespacing*{\section}{0pt}{0.8em}{0.3em}
\titlespacing*{\subsection}{0pt}{0.6em}{0.2em}

\AtBeginEnvironment{thebibliography}{%
    \small
    \setstretch{1.08}
    \setlength{\itemsep}{0.55\baselineskip}
    \setlength{\parskip}{0pt}
    \RaggedRight
}

\setlist[itemize]{itemsep=4pt}
\setlist[enumerate]{itemsep=4pt}

\definecolor{codebackground}{rgb}{0.95,0.95,0.95}
\title{\textbf{PlanCompiler: A Deterministic Compilation\\ 
Architecture for Structured Multi-Step LLM Pipelines}}
\author{
Pranav Harikumar\\
\small Independent Researcher\\
\small \href{https://github.com/prnvh/plancompiler}{github.com/prnvh/plancompiler}
}
\date{}

\begin{document}

\maketitle

\begin{abstract}
Large language models (LLMs) are brittle in multi-step structured workflows, where errors compound across sequential transformations, validation steps, and stateful operations such as SQL persistence. We present PlanCompiler, a deterministic compilation architecture for structured LLM pipelines that separates planning from execution using typed node registries, static graph validation, and topological compilation. Rather than relying on autoregressive chaining at runtime, the system executes a prevalidated workflow graph with explicit parameter and type constraints.

We evaluate the approach on 300 tasks across six benchmark sets covering increasing workflow depth, SQL roundtrip persistence, and schema-themed stress tests. On depth-stratified benchmarks, the compiled system achieves 100\% accuracy on Sets A and B, 88\% on Set C, and 96\% on Set D, outperforming direct code-generation baselines from GPT-4.1 and Claude Sonnet. On Schema trap tasks, it achieves 44/50 compared with 20/50 for GPT-4.1 and 26/50 for Claude. Across the full suite, the compiler consumes approximately \$0.356 in total inference cost, versus \$2.140 for GPT-4.1 and \$18.391 for Claude, while maintaining competitive latency and avoiding the timeout-heavy failure modes observed in free-form baselines.

These results indicate that, in this benchmark setting, deterministic compilation improves first-pass reliability and cost efficiency for registry-constrained structured workflows relative to free-form code generation baselines, while localizing residual failures into two narrow and interpretable classes: late criteria-stage output-contract failures on aggregation tasks and early validation-stage type mismatches at the SQLite persistence boundary.

\end{abstract}


\section{Introduction}
\label{sec:intro}

Large language models have demonstrated substantial capability in generating executable code from natural language descriptions. In single step settings such as producing a sorting function, writing a SQL query, implementing a utility, this capability is well-established and increasingly reliable. The more difficult problem arises when tasks require multiple sequential operations: ingesting structured data, applying a series of transformations, persisting results to a database, and exporting a final artifact. In these multi-step settings, free-form code generation exhibits a characteristic failure mode: errors accumulate across steps. A hallucinated import in step two propagates through steps three and four. A mismatched column name introduced at ingestion causes silent failures at query time. Naming conventions drift between the function a model generates and the function it calls. Each individual step may look locally correct while the composed pipeline fails.

Existing approaches address this in one of two ways. Tool-orchestration frameworks such as LLMCompiler treat the model's output as a DAG of function calls and focus on scheduling and parallel execution, improving latency and cost while accepting execution failure as a runtime concern. NL-to-pipeline synthesis systems such as Text-to-Pipeline define a domain-specific language for data preparation and compile model output to Pandas or SQL, demonstrating that structured intermediate representations improve over direct code generation. Both lines of work normalize the pattern of generating a symbolic representation and executing it deterministically. What is less explicit in this literature is the question of what guarantee the representation boundary provides: which failure modes it eliminates by construction, which it defers to runtime, and which it cannot catch at all.

We describe PlanCompiler, a deterministic compilation architecture that treats this question as its central design concern. The system imposes a strict separation between planning and execution. In the reported benchmark configuration, the LLM's effective role is limited to selecting nodes from a fixed registry and supplying their required parameters, expressed as a typed JSON plan. The planner response format also contains advisory \texttt{flags} and an optional \texttt{glue\_code} override, but \texttt{flags} do not affect validation or compilation and \texttt{glue\_code} is empty in all reported runs. The registry fixes the set of callable primitives; the LLM cannot invent new ones. A deterministic validator then runs seven structural checks on the planâ€”node existence, edge validity, type compatibility, acyclicity, orphan detection, input arity, and required parameter presenceâ€”before any execution occurs. If validation passes, a deterministic compiler assembles an executable Python program from node templates on disk using topological sort. In all reported experiments, execution uses the auto-generated execution block rather than planner-supplied glue code. The LLM is not called again after the plan is emitted. First-pass success under this architecture is the central empirical claim.

All reported results are for ordered single-stream workflows over a fixed node registry; the current implementation does not evaluate true branching, fan-in, repeated node instances, or open-domain tool composition.

We evaluate the system on six benchmark sets totaling 300 tasks, each designed to stress a different aspect of multi-step workflow execution. Sets Aâ€“D vary pipeline depth from short to long compositions, while the remaining sets test robustness under schema-themed task formulations and SQL-mediated state transitions. Across these benchmarks, the compiled system achieves high first-pass success rates and remains substantially more reliable than the free-form code generation baselines.

A secondary finding emerges from the evaluation that appears to reflect a broader design issue rather than an isolated benchmark artifact. Residual PlanCompiler failures localize to two recurring pressure points rather than dispersing across many unrelated mechanisms. One is a late semantic-output boundary in aggregation tasks: grouped aggregation is sometimes routed through QueryEngine SQL instead of the constrained Aggregator path, and the resulting output misses the benchmark's required \texttt{count} column contract. We refer to this as \textit{constraint evasion under partial enforcement}. The other is an early persistence/type boundary failure: after SQLite round-trips, the planner sometimes loses track of the distinction between database handle, file path, and materialized table, and the validator rejects the plan before compilation. We view both as design lessons that may apply more broadly to partially constrained generation systems.

\subsection{Contributions}

The contributions of this paper are three. First, we specify a typed intermediate representation and compiler architecture for LLM program synthesis that eliminates hallucinated imports and many forms of naming and wiring drift by construction rather than by prompting. Second, we present a reliability-focused evaluation protocol with explicit success criteria and a structured failure taxonomy, showing that the architecture maintains high first-pass accuracy across the benchmark sets while free-form generation remains substantially less reliable. Third, we identify and characterize two recurring residual failure families in the current architecture: late output-contract failures on aggregation tasks, often associated with routing work through QueryEngine SQL, and early persistence-boundary type mismatches around SQLite round-trips. We argue that these are design-level concerns rather than isolated prompt-level errors.

\section{Related Work}
\label{sec:related}

This work sits at the intersection of agent orchestration frameworks, natural-language-to-pipeline synthesis, and correctness enforcement for LLM-generated execution. We position PlanCompiler relative to each below.

LangChain and LlamaIndex provide widely used abstractions for building LLM-powered workflows over tools and external components~\citep{langchain, llamaindex}. These frameworks make multi-step composition easier, but they do not impose a hard static correctness boundary before execution. Tool selection and sequencing remain primarily model-driven, and the system can still generate or route arbitrary orchestration logic between calls. In that setting, failures are typically handled at runtime rather than ruled out structurally before execution begins.

Hugging Face's smolagents takes this openness further with an explicitly code-first design~\citep{smolagents}. Its \texttt{CodeAgent} emits executable Python, allowing loops, conditionals, and unrestricted logic between tool calls. While tools themselves may be typed or structured, the orchestration layer remains free-form generated code. PlanCompiler differs along a single architectural axis: the model is not allowed to write execution logic at all. It may only select nodes from a closed registry and supply required parameters. Execution order, node wiring, and runnable code are all derived deterministically after the plan is accepted.

LLMCompiler is structurally close to our system in that it treats tool use as a compiler problem and has the model produce a DAG of function calls for later execution~\citep{kim2023llmcompiler}. The similarity, however, is mainly architectural. LLMCompiler is primarily motivated by efficiency, especially parallel scheduling to reduce latency and cost. In PlanCompiler, plannerâ€“executor separation serves a different purpose: it creates a structural correctness boundary. We are not mainly trying to speed up execution; we are trying to ensure that planning and execution are separated enough for structural guarantees to be enforced before any code runs.

OrchDAG is also closely related~\citep{lu2025orchdag}. It studies complex tool orchestration through DAG-structured plans and includes rule-based verification mechanisms such as structural and dependency checks. That is adjacent to our setting, since both systems rely on structured plans whose internal references can be validated. The difference is where verification sits in the overall design. In OrchDAG, verification primarily supports data generation, evaluation, and training. In PlanCompiler, validation is a runtime enforcement boundary: no plan is compiled unless it passes all seven checks, and that boundary is central to our reliability claims.

Text-to-Pipeline is the closest prior line of work at the architectural level~\citep{ge2025texttopipeline}. Both approaches generate a structured intermediate representation and execute it deterministically rather than emitting unrestricted code directly. The key difference lies in the form of the intermediate representation and in the degree of model freedom. Text-to-Pipeline allows the model to generate DSL programs with substantial expressive flexibility. In the evaluated PlanCompiler configuration, the model solves a constrained selection problem over a closed node registry and cannot invent new primitives; reported experiments use the auto-generated execution path rather than planner-supplied \texttt{glue\_code}. Text-to-Pipeline evaluates functional equivalence between generated and reference pipelines, whereas our evaluation focuses on first-pass task success under increasing structural complexity, with explicit success criteria and a documented failure taxonomy.

AutoPandas also addresses synthesis over a fixed computational surface, using guided search and learned heuristics to construct programs over the pandas API~\citep{bavishi2019autopandas, mckinney2010pandas}. The relationship to our work is clearest at the problem-framing level: both systems generate structured programs over a predefined set of operations. The methods are different, however. AutoPandas is a search-based synthesis system over a broad API surface, whereas PlanCompiler is an LLM-based planner over a much smaller and more tightly constrained registry.

Jigsaw improves generated data-manipulation code through structured post-processing and repair~\citep{jain2022jigsaw}. Its central lesson is that analysis after generation can recover correctness from imperfect outputs. Our approach moves the correction boundary earlier. Rather than repairing invalid outputs after code has already been produced, we reject structurally invalid plans before compilation occurs. The compiler therefore never receives an invalid plan.

Recent work on governed agent execution, including POLARIS, treats agent behavior as typed planning under explicit policy constraints, with validator-guarded execution and strong auditability~\citep{moslemi2026polaris}. This broader direction helps normalize the combination of structured planning, explicit validation, and governed execution in practical deployment settings.

Our contribution relative to this literature is to turn these design patterns into an explicit execution architecture with enforceable structural guarantees. PlanCompiler places a hard validation boundary between planning and execution: the model may propose a plan, but no code is generated or run unless that plan satisfies a closed set of structural checks. The empirical evaluation is intended to test the architectureâ€™s boundary conditions, showing which failure modes disappear under this boundary and which persist. In particular, the QueryEngine evasion result---where partial enforcement causes the model to route aggregation work into a semantically open SQL surface---identifies one practically important weakness that may arise in partially constrained generation systems.

\section{Method}
\label{sec:method}

\subsection{\textbf{System Architecture}}

PlanCompiler consists of five components arranged in a strictly ordered pipeline. Only one component involves a language model call. All subsequent system logic is non-generative; validation and compilation are deterministic, and execution is deterministic given the emitted program, fixtures, and environment.

The planner receives a natural-language task description together with the typed node registry and emits a JSON plan specifying node selections, parameter bindings, and edges. This plan is then passed to a static validator that checks structural correctness before any code is generated. Only validated plans are compiled into executable Python assembled from node templates on disk.

\subsection{\textbf{Graph Specification}}

A node is a typed computational primitive defined as a Pydantic model with seven fields: name, description, input\_type, output\_type, template\_path, required\_params, and function\_name. Every field is fixed at registry definition time and left untouched at runtime. The function\_name field is stored explicitly â€” the compiler uses it directly, with no string manipulation, snake\_case conversion, or regex derivation from the node name. This is a deliberate design choice: the mapping from node to callable is a registry fact, not a naming convention. A separate template\_path field points to a pre-written Python file on disk. The compiler reads this file verbatim; the LLM never sees template contents.

An edge is a directed pair [source, target] asserting that the output of source is passed as the first positional argument to target. Edges are the only mechanism for data flow between nodes â€” there is no shared mutable in-memory state, no global context, no implicit passing. In the current auto-generated execution path, each node may consume at most one predecessor, so the operational subset is a single-stream pipeline rather than a general fan-in DAG. The plan schema uses node names as unique identifiers, meaning a given node type can appear at most once per plan. Repeated use of the same primitive â€” for example, two sort stages or two instances of the same database node type â€” is structurally inexpressible in the current schema; this is a documented architectural constraint.

Parameters are key-value pairs attached to a node in the plan's parameters dict. They become keyword arguments in the compiled execution call. Parameters are LLM-supplied; the registry declares which parameters are required, and the validator enforces their presence. The compiler emits parameter values using repr(), so planner-supplied values are inserted directly into the emitted code as Python literals rather than being regenerated or transformed by a second model step. Benchmark-critical parameters are expressed via \texttt{required\_params} in the registry; some templates still accept optional helper arguments with defaults, but these are not validator-enforced and are not central to the reported benchmark protocol. The DataSorter case during development showed that optional parameters with defaults can cause silent wrong-output failures: a task specifying descending sort would pass file\_has\_column criteria while producing incorrectly ordered output. Making ascending a required parameter forces the validator to catch omissions rather than letting them silently produce incorrect output.

The type system defines five types: FilePath, DataFrame, DBHandle, HTTPResponse, and ANY, implemented as a string enum for registry serialization and validation. An edge is type-compatible if source.output\_type == target.input\_type, or if either side is ANY. The ANY wildcard is exercised by Logger and is also used in the registry-level typing of ErrorHandler. Type compatibility is checked statically before any execution begins. A type mismatch is a hard abort, not a warning.

\subsection{\textbf{Planner}}

The planner is the only component that involves a language model call. It receives a natural language task description alongside a serialized representation of the node registry â€” names, descriptions, input/output types, and required parameters â€” and returns a normalized JSON plan. The planner response format includes five top-level fields: nodes (an ordered list of registry node names), edges (directed pairs defining data flow), parameters (key-value pairs per node), flags (advisory metadata such as missing-node or credential warnings, ignored by validation and compilation), and glue\_code (an optional execution-block override, empty in all reported benchmark runs).

The planner includes a normalization step that handles non-standard LLM output formats before the plan reaches the validator: nodes returned as dicts with type and params fields, edges returned as dicts with from and to fields, edges returned as string arrow notation, integer node references, and node names in snake\_case rather than CamelCase are all normalized to the canonical schema. For the reported benchmark setting, the harness also applies a pre-validation edge normalization: if the emitted edge list is malformed, incomplete, or non-forward, it is replaced with the ordered chain node[i] $\rightarrow$ node[i+1]. The validator therefore always receives a structurally normalized single-stream plan in these experiments.

\subsection{\textbf{Validation Layer}}

The validator is the primary explicit rejection boundary for plans. If a plan passes all seven checks, it is guaranteed to satisfy the compilerâ€™s structural preconditions. There is no partial validation, no warnings-only mode, and no retry. Seven checks run in strict order, with early abort on CHECK 1 because subsequent checks assume node existence.

CHECK 1 (node existence) binds the plan to the registry: any node name not present in NODE\_REGISTRY is rejected immediately. CHECK 2 (edge validity) confirms that every node referenced in an edge is declared in the plan's node list. CHECK 3 (type compatibility) compares output and input types across every edge, with the ANY wildcard exempting either direction. CHECK 4 (acyclicity) runs a topological sort via Kahn's algorithm; if not all nodes are visited, a cycle exists and the plan is rejected. CHECK 5 (no orphans) ensures every node in a multi-node plan appears in at least one edge. CHECK 6 (input arity) enforces that every non-entry-point node has exactly one inbound edge, ruling out fan-in structures the current execution model does not support. This is the key reason the current shipped implementation supports single-stream pipelines rather than true multi-input execution. CHECK 7 (required parameters) verifies the presence of every required parameter for every node, checking presence rather than validity â€” the validator confirms that condition is present for DataFilter but does not evaluate whether the expression is a valid pandas query string.

The validator exposes topological\_sort() as a public function used directly by the compiler. This is the only shared interface between the two components.

\subsection{\textbf{Compilation}}

The compilation stage in PlanCompiler is called only after the validator returns clean. It re-runs validate\_plan internally as a defensive check â€” validation occurs twice, and the compiler does not trust that callers have pre-validated. Topological sort via Kahn's algorithm produces a deterministic linear execution order from the DAG. For the linear pipelines that constitute all current benchmark tasks, this is simply the node list in dependency order.

Template assembly proceeds as follows: for each node in topological order, the compiler reads the corresponding template file from disk verbatim and appends it to the output buffer. There is no string manipulation of template content and no interpolation. The execution block is auto-generated by default, and all reported benchmark runs use this path: the compiler iterates the topological order, looks up each node's function\_name from the registry directly, identifies the predecessor via the edges map, and emits code of the form out\_<function\_name> = <function\_name>(out\_<predecessor>, key=value, ...), where planner-supplied parameter values are inserted via \texttt{repr()}. The last node's output is printed.

\subsection{\textbf{Execution Model}}

Execution is a single Python subprocess invocation of the compiled app.py. There is no intermediate interpreter, virtual machine, or runtime planning layer between the emitted program and execution. Inter-node execution state passes through function return values: each node function takes the previous node's output as its first positional argument and returns a new value. In the auto-generated execution path used in the benchmark, exactly one upstream value is threaded into each non-entry node. There is no shared mutable in-memory state between nodes and no implicit side-channel communication. Once app.py is emitted, the LLM is not called again. There is no output inspection loop, no repair prompt, and no autoregressive chaining at runtime. If execution fails, it fails. The reliability claim is about first-pass success, not success after correction. Adding a repair loop would likely reduce non-runtime failures, but it would change the claim being evaluated: the present results measure first-pass execution success, not success after iterative correction.

Determinism is a property of the compilation stage rather than the full end-to-end system: given the same valid plan and the same input fixtures, the compiler always emits identical \texttt{app.py} content, verifiable by output hash. The remaining source of non-determinism is the planner itself, which may produce different valid plans across runs. By contrast, free-form baseline generation showed small but measurable multi-run variance even at temperature=0; we analyze this in Section~\ref{sec:results}.

\begin{figure}[H]
    \centering
    \includegraphics[width=0.5\linewidth]{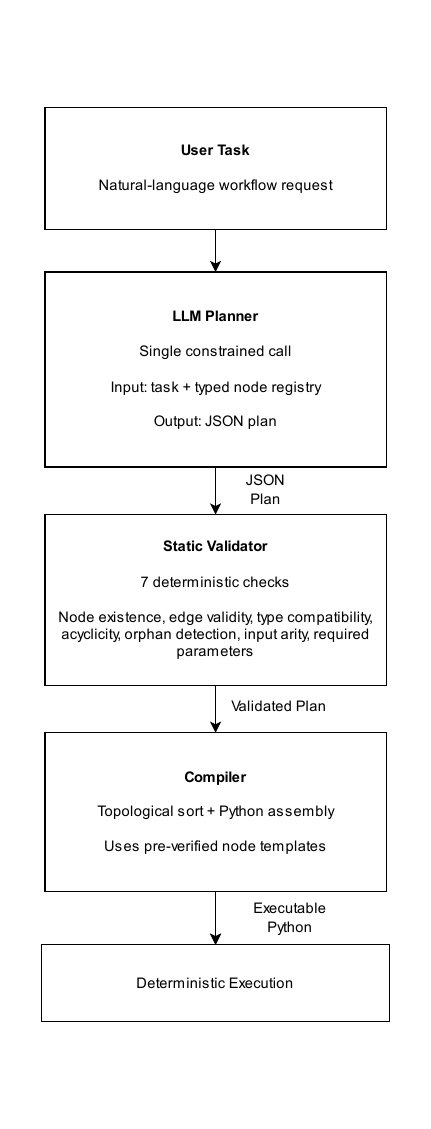}
    \caption{Overview of PlanCompiler. A single LLM call produces a structured JSON plan over a typed node registry. The plan is validated using deterministic structural checks before compilation. Only validated plans are compiled into executable Python and run.}
    \label{fig:architecture}
\end{figure}

\section{Experimental Setup}
\label{sec:experiments}

\subsection{\textbf{Benchmark Design}}

The benchmark measures first-pass success rate: the fraction of tasks where the full pipeline â€” plan, validate, compile, execute, criteria check â€” succeeds end-to-end with zero human intervention. A task is scored as a success only if every criterion passes on the first attempt. There is no retry and no partial credit.

For each task, the harness tracks: (1) \textbf{success/failure} against all criteria, (2) \textbf{planner latency and token usage} (the planner uses GPT-4o-mini at temperature=0, measuring input/output tokens and USD cost), (3) \textbf{end-to-end wall-clock latency} from plan to execution completion, and (4) \textbf{compilation success} (whether generated code is executable). Per-task measurements enable detailed failure analysis across pipeline stages.

Tasks are organized into six sets totaling 300 tasks. Sets A through D approximate a complexity ladder stratified primarily by pipeline length. Sets E and F are specialized stress tests targeting distinct failure modes.

\begin{center}
\begin{tabular}{lllp{8cm}}
\toprule
\textbf{Set} & \textbf{Depth} & \textbf{Count} & \textbf{Focus} \\
\midrule
A & 3--5 nodes & 50 & Shallow pipelines. Ingestion, single transforms, export. \\
B & 5--8 nodes & 50 & Medium pipelines. Chained transforms, SQLite roundtrips, aggregations. \\
C & 8--10 nodes & 50 & Long pipelines. Multi-step SQLite, post-query transforms, complex aggregations. \\
D & 10+ nodes & 50 & Maximum complexity within library scope. \\
E & Mixed & 50 & Schema-themed tasks. Null handling, type casting and aggregation-output contracts. \\
F & Mixed & 50 & SQL roundtrip tasks. CSV/JSON $\rightarrow$ SQLite $\rightarrow$ export data integrity. \\
\bottomrule
\end{tabular}
\end{center}

Each set is stored as a fixed 50-task JSON file. The shipped task files do not carry an explicit original/probe annotation, so results are reported at the set level rather than stratified by such a split.

All planning operations run at temperature=0 to reduce sampling variance in planner output. Token counts and latency measure pure planning cost without retry or correction loops.

The current benchmark is intentionally aligned with the present PlanCompiler library and application domain, and the reported results should therefore be understood as evidence for this class of structured data-processing workflows rather than as a blanket claim about arbitrary code-generation problems. We compare primarily against free-form code generation because it remains the standard baseline in contemporary LLM code-generation practice, and thus provides the most meaningful reference point for testing whether a constrained planning-and-compilation architecture improves reliability.

\subsection{\textbf{Success Criteria}}

The benchmark harness supports five criterion types, evaluated after each task execution. All criteria for a task must pass; if any criterion fails, the task is marked as failed.

\begin{center}
\begin{tabular}{lp{10cm}}
\toprule
\textbf{Criterion} & \textbf{Check} \\
\midrule
file\_exists & Output file exists at the specified relative path (from run\_dir). The current checker uses \texttt{os.path.exists} and therefore does not impose any additional case-sensitivity rule beyond the host filesystem. \\
file\_has\_column & Tabular output (CSV or JSON) contains a column with the exact name specified. Column name matching is case-sensitive. \\
file\_row\_count & Tabular output contains exactly the specified number of rows (determined by loading the file and calling len(df)). \\
stdout\_contains & Process stdout contains the specified substring (exact match, case-sensitive, no regex). \\
file\_column\_sorted & Named column is monotonically sorted in the specified direction (ascending or descending). NaN values are excluded from the sort check; ties are allowed (i.e., non-strict monotonic ordering). \\
\bottomrule
\end{tabular}
\end{center}

file\_column\_sorted was added to prevent sort-direction correctness bugs. Prior to this, tasks specifying sort direction were under-verified: a pipeline that sorted ascending when descending was required would pass file\_has\_column criteria while producing incorrect output. This criterion type is required for any task where sort direction is part of the correctness specification. 

\subsection{Baseline}

The baseline generates free-form Python directly from the task description, with no scaffolding, no node registry, and no structural constraints. Each baseline model receives the same natural language task description as the compiler's planner. Each task is run N\_RUNS=3 times, and success requires all three runs to pass, matching the compiler's evaluation protocol. Baseline programs run in a separate harness with a hard 40-second timeout per task; tasks requiring long-running processes such as Flask servers are handled with process group termination to prevent indefinite hang.

We evaluate two baseline models: \textbf{GPT-4.1} and \textbf{Claude Sonnet 4.6}, both at temperature=0 to reduce sampling variance. These models represent different architectural and training approaches, providing a broader baseline comparison. GPT-4.1 was used instead of GPT-4o because preliminary testing showed materially worse SQLAlchemy-version compatibility in our environment.

Each baseline model is evaluated on all 300 tasks (50 per set, 3 runs per task). Results are reported separately for each model and set in Section~\ref{sec:results}.

\subsection{Evaluation Metrics}

We track metrics across two dimensions: per-task granularity and per-set aggregation. All metrics are computed from three runs (N\_RUNS=3) of each task, enabling analysis of both reliability and consistency. 

\textbf{Per-Task Metrics}

\begin{center}
\begin{tabular}{lp{9cm}}
\toprule
\textbf{Metric} & \textbf{Description} \\
\midrule
plan\_success & LLM returned a valid, parseable JSON plan. \\
validation\_success & Plan passed all seven validator checks (node existence, edge validity, type compatibility, acyclicity, orphan detection, input arity, required parameters). \\
compile\_success & Compiler emitted a valid, syntactically correct Python artifact. \\
run\_success & Artifact executed with exit code 0 (no runtime error). \\
criteria\_passed & Output matched all task success criteria. \\
first\_pass\_success & All five stages succeeded on all three runs without needing repair. \textit{Primary compiler metric}. \\
pass\_count & Number of runs (out of 3) where all five stages succeeded. \\
avg\_duration\_seconds & Mean wall-clock latency across three runs (plan through execution completion). \\
planner\_input\_tokens & Total input tokens consumed by planner across three runs. \\
planner\_output\_tokens & Total output tokens generated by planner across three runs. \\
planner\_cost\_usd & USD cost of planning (GPT-4o-mini pricing at \$0.15/1M input, \$0.60/1M output). \\
\bottomrule
\end{tabular}
\end{center}

Baseline models (GPT-4.1 and Claude Sonnet 4.6) report equivalent metrics with their own prefixes (e.g., baseline\_gpt41\_first\_pass\_success, baseline\_claude\_first\_pass\_success). Each baseline model is evaluated independently on the same tasks and success criteria. 

\textbf{Per-Set Aggregation}

For each set of 50 tasks, we compute:

\begin{center}
\begin{tabular}{lp{9cm}}
\toprule
\textbf{Metric} & \textbf{Definition} \\
\midrule
Success Rate (\%) & Fraction of 50 tasks where first\_pass\_success=true. \\
Average Latency (s) & Mean of avg\_duration\_seconds across 50 tasks. \\
Total Cost (USD) & Sum of planner\_cost\_usd across 50 tasks. \\
Average Cost per Task (USD) & Total cost / 50. \\
Average Input Tokens & Mean of planner\_input\_tokens across 50 tasks. \\
Average Output Tokens & Mean of planner\_output\_tokens across 50 tasks. \\
\bottomrule
\end{tabular}
\end{center}

\textbf{Comparison Protocol}

The primary comparison is \textbf{first\_pass\_success} (compiler) versus \textbf{first\_pass\_success} (baseline) aggregated per set and across all 300 tasks.Â 

Here, first-pass success means success without any iterative correction, repair, or prompt alteration. To test whether this property holds consistently rather than accidentally, both baseline and compilerÂ  tasks are executed three times under identical conditions, and success is recorded only when all three first-pass runs succeed.

We report:
\begin{enumerate}
    \item \textbf{Per-set success rates} for compiler, GPT-4.1, and Claude Sonnet 4.6
    \item \textbf{Overall success rate} (aggregate across all six sets)
    \item \textbf{Cost per successful task} (total planner cost / number of tasks passing criteria)
    \item \textbf{Latency} (mean execution time, excluding tasks that timed out)
\end{enumerate}
This enables analysis of: (a) how accuracy scales with pipeline depth (Sets A-D), (b) whether specialization stress tests (Sets E-F) expose systematic weaknesses, and (c) whether the compiler's reliability gains come at the cost of increased latency or planning expense.

\section{Results}
\label{sec:results}

\subsection{Main Results}

Table~\ref{tab:success_rates} reports first-pass success rates for the compiled system, GPT-4.1 baseline, and Claude Sonnet 4.6 baseline across all six benchmark sets (n=50 per set, n=300 total).

\begin{table}[H]
\centering
\caption{First-pass success rates across all benchmark sets.}
\label{tab:success_rates}
\small
\begin{tabularx}{\textwidth}{llllll}
\toprule
\textbf{Set} & \textbf{Focus} & \textbf{PlanCompiler} & \textbf{GPT-4.1} & \textbf{Claude 4.6} & \textbf{$\Delta$ vs GPT-4.1} \\
\midrule
A & 3--5 nodes & 50/50 (100\%) & 38/50 (76\%) & 30/50 (60\%) & +24 pts \\
B & 5--8 nodes & 50/50 (100\%) & 36/50 (72\%) & 23/50 (46\%) & +28 pts \\
C & 8--10 nodes & 44/50 (88\%) & 34/50 (68\%) & 27/50 (54\%) & +20 pts \\
D & 10+ nodes & 48/50 (96\%) & 38/50 (76\%) & 36/50 (72\%) & +20 pts \\
E & Schema traps & 44/50 (88\%) & 20/50 (40\%) & 26/50 (52\%) & +48 pts \\
F & SQL roundtrips & 42/50 (84\%) & 36/50 (72\%) & 45/50 (90\%) & +12 pts \\
\midrule
\textbf{Overall} & \textbf{All 300} & \textbf{278/300 (92.67\%)} & \textbf{202/300 (67\%)} & \textbf{187/300 (62\%)} & \textbf{+25.67 pts} \\
\bottomrule
\end{tabularx}
\end{table}

PlanCompiler achieves 84--100\% success across all sets, with perfect 100\% on Sets A and B and 96\% on Set D. The compiler's advantage over GPT-4.1 ranges from 12 points (Set F) to 48 points (Set E), with an aggregate delta of 25.67 percentage points (92.67\% vs 67\%).

Claude Sonnet 4.6 exhibits different failure patterns: it achieves 90\% on Set F (SQL roundtrips)---the only system to exceed the compiler on a set---but falls to 46\% on Set B and 52\% on Set E. The compiler's overall advantage over Claude is 30.33 percentage points (92.67\% vs 62.33\%).

\subsection{Token Cost Analysis}

Planning cost (GPT-4o-mini for compiler planning) and execution cost (GPT-4.1 and Claude for baseline generation) are reported per task and aggregated across all 300 tasks.

\subsubsection{Cost Summary}

\begin{table}[H]
\centering
\caption{Cost efficiency across all models.}
\label{tab:cost_summary}
\small
\begin{tabularx}{\textwidth}{llllll}
\toprule
\textbf{Model} & \textbf{Total Cost} & \textbf{Successes} & \textbf{Cost/Success} & \textbf{Mean Cost/Task} \\
\midrule
PlanCompiler & \$0.3562 & 278 (92.67\%) & \$0.001281 & \$0.001187 \\
GPT-4.1 & \$2.1396 & 202 (67\%) & \$0.010592 & \$0.007132 \\
Claude Sonnet & \$18.3907 & 187 (62\%) & \$0.098346 & \$0.061304 \\
\bottomrule
\end{tabularx}
\end{table}

The compiler is 8.27x more cost-efficient than GPT-4.1 and 76.75x more efficient than Claude on a cost-per-successful-task basis. Claude's dramatically higher cost (\$18.39 vs \$0.36) is driven by consistently higher output token generation across all sets, even on failed tasks.

Although PlanCompiler uses GPT-4o-mini for planning, we do not regard the use of a cheaper model solely as a caveat; rather, it is part of the architectural result. A central benefit of the system is that by narrowing the modelâ€™s role to structured planning over a typed registry and shifting correctness burden into validation and compilation, it can obtain strong end-to-end performance with a smaller, lower-cost model.

\subsubsection{Per-Set Token Usage}

\begin{table}[H]
\centering
\caption{Per-set token metrics (mean values per task).}
\label{tab:tokens_per_set}
\small
\begin{tabularx}{\textwidth}{lllllll}
\toprule
\textbf{Set} & \textbf{PC Input} & \textbf{PC Output} & \textbf{GPT Input} & \textbf{GPT Output} & \textbf{Claude Input} & \textbf{Claude Output} \\
\midrule
A & 3,190 & 608 & 301 & 429 & 331 & 3,325 \\
B & 3,240 & 991 & 351 & 584 & 384 & 4,022 \\
C & 3,318 & 1,482 & 429 & 894 & 467 & 4,721 \\
D & 3,360 & 1,777 & 471 & 1,231 & 513 & 5,790 \\
E & 3,234 & 967 & 345 & 898 & 377 & 3,810 \\
F & 3,285 & 1,143 & 396 & 739 & 440 & 2,351 \\
\bottomrule
\end{tabularx}
\end{table}

\textbf{Key observations:}

\begin{enumerate}
    \item PlanCompiler planning tokens scale with pipeline complexity: Input tokens increase from 3,190 (Set A, simple) to 3,360 (Set D, complex). Output tokens increase from 608 to 1,777, reflecting the planning difficulty of longer pipelines.
    
    \item Baseline execution tokens show high variance: GPT-4.1 output (429$\rightarrow$1,231) stays bounded. Claude output (3,325$\rightarrow$5,790) is 5-10x higher than input, indicating verbose code generation with significant redundancy.
    
    \item Claude generates excessive output: Even on simple tasks (Set A), Claude generates 3,325 tokens while GPT-4.1 generates 429 tokens for the same input. This verbosity persists across all sets. Its relationship to Claude's failure modes is analyzed in Section~\ref{sec:failures}.
\end{enumerate}

\subsubsection{Cost Scaling with Pipeline Depth}

\begin{table}[H]
\centering
\caption{Compiler cost per task increases with pipeline depth; success rates remain high.}
\label{tab:cost_depth}
\small
\begin{tabularx}{\textwidth}{lllll}
\toprule
\textbf{Depth} & \textbf{Task Count} & \textbf{PC Cost} & \textbf{PC Success} & \textbf{Cost/Success} \\
\midrule
3--5 nodes & 50 & \$0.04 & 100\% & \$0.00084 \\
5--8 nodes & 50 & \$0.05 & 100\% & \$0.00108 \\
8--10 nodes & 50 & \$0.07 & 88\% & \$0.00157 \\
10+ nodes & 50 & \$0.08 & 96\% & \$0.00164 \\
Mixed (E, F) & 100 & \$0.11 & 86\% & \$0.001304 \\
\bottomrule
\end{tabularx}
\end{table}

PlanCompiler incurs higher planning cost on deeper pipelines while maintaining high success rates. Deeper pipelines cost more to plan while maintaining high overall accuracy. Baselines do not follow this pattern---they pay similar execution costs (\$0.0040-0.0884/task) but with highly variable success rates.

\textbf{Summary}

The compiler's planning investment (\$0.00119/task on average) is recovered through superior reliability. Cost per successful task is 8.4x lower than GPT-4.1 and 77.8x lower than Claude. Baseline models, despite appearing cheaper on a per-task basis, fail frequently, making their true cost-per-reliable-result substantially higher. The compiler trades higher planning cost for lower failure cost, yielding net savings in benchmark settings where first-pass correctness is important.

\subsection{Latency Analysis}

Wall-clock latency measures end-to-end time from task submission to success/failure determination, including plan generation, compilation, and execution.

\begin{table}[H]
\centering
\caption{Mean latency per task (seconds).}
\label{tab:latency}
\small
\begin{tabularx}{\textwidth}{llll}
\toprule
\textbf{Set} & \textbf{PlanCompiler} & \textbf{GPT-4.1} & \textbf{Claude} \\
\midrule
A & 5.79s & 3.57s & 14.63s \\
B & 8.54s & 4.30s & 16.67s \\
C & 11.72s & 5.81s & 17.26s \\
D & 12.05s & 5.88s & 20.75s \\
E & 8.50s & 5.64s & 16.47s \\
F & 9.64s & 4.91s & 11.32s \\
\midrule
\textbf{Mean} & \textbf{9.37s} & \textbf{5.05s} & \textbf{16.15s} \\
\bottomrule
\end{tabularx}
\end{table}

\textbf{Key findings:}

\begin{enumerate}
    \item PlanCompiler latency increases monotonically with pipeline complexity (5.79s $\rightarrow$ 12.05s, Sets A$\rightarrow$D). This is expected: longer pipelines require longer planning and more complex code generation.
    
    \item Baseline latency is lower on average but decoupled from success: GPT-4.1 averages 5.05s despite only 67\% success rate. Claude averages 16.15s despite 62\% success. 
\end{enumerate}

\subsection{Baseline Non-Determinism}
\label{sec:nondeterminism}

The GPT-4.1 baseline at temperature=0 showed measurable multi-run variance on repeated executions of the same tasks. On Set F, the first run achieved 42/50 success while the second run achieved 38/50---an 8\% variance from identical prompts and temperature. 

This provides a small but concrete example of the run-to-run variance that the architecture is designed to reduce. Even temperature=0 baseline generation exhibits multi-run variance, while deterministic compilation is reproducible by construction.

\subsection{Constraint Evasion}
\label{sec:evasion}

The compiled system's failures are not randomly distributed. Across the benchmark sets where the system falls below perfect accuracy---especially C, E, and F---a systematic failure pattern emerges.

QueryEngine exposes an unconstrained raw SQL string parameter. More broadly, the validator enforces required-parameter presence rather than semantic validity; for example, \texttt{DataFilter.condition} is also planner-supplied and semantically unchecked. The planner response schema also contains an unchecked \texttt{glue\_code} field, but it is empty in all reported benchmark runs. In tasks requiring grouped aggregation, the planner systematically attempts to satisfy grouped aggregation tasks by embedding GROUP BY and COUNT(*) in the QueryEngine SQL string rather than routing the computation through the Aggregator node.

For example:

\textbf{Expected:} Aggregator node with \texttt{agg\_func: ``count''} producing column \texttt{count}

\textbf{Generated:} QueryEngine with SQL \texttt{SELECT group\_col, COUNT(*) FROM table GROUP BY group\_col}

The pipeline executes without error (the SQL is valid), but output validation fails because the output does not expose the exact required column name \texttt{count}.

\begin{table}[H]
\centering
\caption{Constraint evasion accounts for 13 of 22 total PlanCompiler failures (59.1\%).}
\label{tab:evasion}
\small
\begin{tabularx}{\textwidth}{lllll}
\toprule
\textbf{Set} & \textbf{Total Failures} & \textbf{QueryEngine Evasion} & \textbf{Other Failures} & \textbf{Evasion \%} \\
\midrule
A & 0 & 0 & 0 & --- \\
B & 0 & 0 & 0 & --- \\
C & 6 & 5 & 1 & 83\% \\
D & 2 & 1 & 1 & 50\% \\
E & 6 & 4 & 2 & 67\% \\
F & 8 & 3 & 5 & 38\% \\
\midrule
\textbf{Total} & \textbf{22} & \textbf{13} & \textbf{9} & \textbf{59\%} \\
\bottomrule
\end{tabularx}
\end{table}

\textbf{Interpretation:} This failure family is better explained as a consequence of partial semantic constraint enforcement than as a generic hallucination failure. In these tasks, the planner generates structurally valid and executable pipelines, but routes grouped aggregation through a semantically open SQL surface rather than the constrained Aggregator path. This pattern is most prominent in Sets C and E, and remains present in Sets D and F, suggesting that aggregation-heavy tasks make this routing behavior more likely without exhausting the full residual error story.

\section{Failure Analysis}
\label{sec:failures}

\subsection{Overview}

Across 300 benchmark tasks, PlanCompiler produced 22 failures, GPT-4.1 produced 98 failures, and Claude Sonnet 4.6 produced 113 failures. PlanCompiler failures are concentrated in two narrow patterns: 13 tasks fail only at the final criteria stage after successful planning, validation, compilation, and execution, while 9 tasks are rejected earlier by static validation due to type-boundary errors around SQLite persistence. Because each task is executed three times, task-level failure counts and run-level failed attempts are not identical; run counts below refer to failed runs across the repeated-execution protocol. Baseline failures are broadly distributed across error types that compound with pipeline length and task specification clarity.

\subsection{PlanCompiler Failure Modes}

\textbf{Pattern 1: QueryEngine Evasion (13 failures, 59.1\% of PlanCompiler failures)}

The planner satisfies aggregation tasks by embedding GROUP BY and COUNT(*) in the raw SQL string passed to QueryEngine, rather than constructing a pipeline that routes through the Aggregator node. The resulting output does not expose the exact required column name \texttt{count}, failing \texttt{file\_has\_column} criteria. The compiled artifact executes without error---exit code 0, no exceptions---making this failure invisible to \texttt{run\_success} and only detectable at the criteria evaluation stage.

The planner is not hallucinating or producing structurally invalid plans in these cases. It is producing valid, executable pipelines that miss the benchmark's required output contract by routing grouped aggregation through a semantically open SQL surface rather than through the constrained Aggregator path. This makes the failure architectural in an important sense, but it is not the sole residual failure mode in the system.

Evasion instances are distributed unevenly across sets, suggesting that aggregation-heavy tasks are more prone to this routing behavior than simpler ones.

\textbf{Pattern 2: DBHandle Type-Confusion Errors (9 failures, 40.9\% of PlanCompiler failures)}

The remaining nine failures are validator-detected \texttt{TYPE\_MISMATCH}s concentrated around DBHandle flow after \texttt{SQLiteConnector}. These are not semantically subtle, validator-clean graphs; they are structurally invalid plans rejected before compilation.

In seven tasks, the planner places \texttt{SQLiteReader} after \texttt{SQLiteConnector}, yielding \texttt{TYPE\_MISMATCH: [SQLiteConnector -> SQLiteReader] NodeType.DB\_HANDLE != NodeType.FILE\_PATH}. One of those seven then also routes \texttt{SQLiteReader} into \texttt{DataSorter}, adding \texttt{TYPE\_MISMATCH: [SQLiteReader -> DataSorter] NodeType.DB\_HANDLE != NodeType.DATA\_FRAME}. The remaining two tasks wire \texttt{SQLiteConnector} directly into \texttt{Aggregator} and \texttt{CSVExporter}, both of which require \texttt{DataFrame} input. This family contributes nine failed tasks across sets C=1, D=1, E=2, and F=5, corresponding to 12 failed runs under the repeated-execution protocol. Taken together, these failures indicate planner confusion at the persistence boundary: the distinction between database handle, database path, and materialized table is lost, and the validator catches the mismatch cleanly.

\subsection{Baseline Failure Modes}

\subsubsection{GPT-4.1 Failures (98 failures, 32.7\% of tasks)}

GPT-4.1 failures are distributed across five recurring error types. The dominant failure mode is aggregation column naming (\textasciitilde45\% of failures), followed by schema drift and hallucinated columns (\textasciitilde20\%).

\begin{center}
\begin{tabular}{lp{10cm}}
\toprule
\textbf{Failure Type} & \textbf{Description} \\
\midrule
Aggregation column naming & \textasciitilde45\% Count column named occurrences, frequency, total, or num\_rows instead of count. Fails file\_has\_column criteria. \\
Schema drift / hallucinated columns & \textasciitilde20\% GPT-4.1 generates rigid schema expectations (e.g., Expected column: OrderID) not present in the fixture. Downstream validation failures. \\
Deprecated API patterns & \textasciitilde15\% SQLAlchemy 1.4 engine.execute() pattern generated for SQLAlchemy 2.0 environment. Requires compatibility layer; residual failures in deeply nested calls. \\
Import errors / hallucinated libraries & \textasciitilde12\% Imports of non-existent modules or incorrect function signatures from real modules. Fails at run\_success with ModuleNotFoundError or AttributeError. \\
Structural drift across pipeline length & \textasciitilde8\% Variable naming inconsistency between pipeline steps---a DataFrame produced in step 2 referenced under a different name in step 4. Fails silently or raises NameError. \\
\bottomrule
\end{tabular}
\end{center}

The dominant baseline failure mode---free-form aggregation column naming drift inside generated code---does not appear in the same form in compiler results. Instead, related output-contract failures appear only when aggregation is routed through QueryEngine rather than through the constrained Aggregator node. This is the clearest concrete demonstration of what the constraint architecture provides: when the node implementation is fixed, the model cannot arbitrarily rename aggregation outputs inside generated code.

Schema drift and hallucinated column expectations illustrate a related failure mode. GPT-4.1 generates code that asserts the presence of specific column names not guaranteed by the fixture, then fails when those columns are absent or named differently. The compiler avoids much of this specific failure mode because its shipped templates generally operate over observed columns without synthesizing new schema expectations. 

\subsubsection{Claude Sonnet 4.6 Failures (113 failures, 37.7\% of tasks)}

Claudeâ€™s failures are often associated with verbose code generation and additional assumptions not clearly required by the task description. Unlike GPT-4.1's systematic column-naming issues, Claude's failures stem from verbose code generation that adds features the task does not require.

\textbf{Token-Based Over-Engineering Pattern:}

Across all sets, failed Claude runs consume significantly more output tokens than passing runs:

\begin{center}
\begin{tabular}{lllll}
\toprule
\textbf{Set} & \textbf{Pass Avg Output Tokens} & \textbf{Fail Avg Output Tokens} & \textbf{Ratio} \\
\midrule
A & 996.3 & 1416.2 & 1.42x \\
B & 1221.8 & 1540.0 & 1.26x \\
C & 1358.7 & 2059.6 & 1.52x \\
D & 1877.2 & 2150.7 & 1.15x \\
E & 880.7 & 1853.9 & 2.10x \\
F & 666.8 & 2013.8 & 3.02x \\
\bottomrule
\end{tabular}
\end{center}

Failed runs use 1.15--3.02x more output tokens than passing runs. This pattern is consistent with Claude's over-engineering tendency: when given sparse task descriptions, failed runs are often more verbose rather than more targeted.

\textbf{Specific Failure Patterns:}

\begin{enumerate}
    \item \textbf{JSON Input Handling:} Claude can generate code that outputs raw Python objects instead of proper JSON, causing \texttt{[ERROR] Top-level JSON must be an object.} errors. In the saved results, this exact error appears once, on \texttt{set\_a\_29\_products\_validate\_export} (run 3), across the benchmark's 99 JSON-input tasks.
    
    \item \textbf{Schema Validation Tasks:} Tasks explicitly requesting schema validation trigger a consistent failure mode: Claude generates validators that raise exceptions on the actual fixture data, treating validation as ``raise if data is wrong'' rather than ``validate and allow.'' Schema validation task density is especially high in Set E (92\%, versus 6\% in Set A), and this coincides with elevated Claude failure rates on that set.
    
    \item \textbf{Logger-Tagged Tasks:} In the saved results, Claude fails on logger-tagged tasks at counts A=2, B=0, C=5, D=7, E=5, and F=3. These failures do not reduce to a single stdout/stderr-mixing mechanism: the observed errors include schema-validation exceptions, timeouts, and at least one logging-related Unicode encoding failure on Windows.
\end{enumerate}

\subsection{Failure Mode Comparison}

The structural difference between PlanCompiler and baseline failure distributions reflects the architectural difference between the two systems.

Baseline failures are broad because free-form generation has many surfaces on which errors can occur and no mechanism to catch them before execution. GPT-4.1 and Claude generate different wrong outputs for the same task (different column names, different assumptions about data format), but both are systematic in their wrongness.

PlanCompiler failures are narrow because the architecture eliminates many free-form error surfaces by construction and localizes the remaining errors into two interpretable families rather than distributing them across many unrelated mechanisms. The larger of these families involves grouped aggregation routed through QueryEngine's semantically open SQL surface, accounting for 13 of 22 failed tasks (59.1\%). The second involves DBHandle type confusion at the SQLite persistence boundary, accounting for the remaining 9 failed tasks (40.9\%). Closing the QueryEngine surface would likely improve reliability further, but it would not by itself eliminate all observed residual failures.

In contrast, baseline failures are spread across five qualitatively different error types (column naming, schema drift, deprecated API patterns, imports, structural inconsistency), indicating no clear path to improvement through single-surface fixes. Improving baseline reliability would require addressing all error surfaces simultaneously.

\subsection{What Validation Catches and Does Not Catch}

\begin{center}
\begin{tabular}{lll}
\toprule
\textbf{Failure Mode} & \textbf{Caught By} & \textbf{Stage} \\
\midrule
Hallucinated node name & CHECK 1 (node existence) & Pre-compilation \\
Invalid edge reference & CHECK 2 (edge validity) & Pre-compilation \\
Type mismatch between nodes & CHECK 3 (type compatibility) & Pre-compilation \\
Cyclic dependency & CHECK 4 (acyclicity) & Pre-compilation \\
Disconnected node & CHECK 5 (no orphans) & Pre-compilation \\
Fan-in wiring & CHECK 6 (input arity) & Pre-compilation \\
Missing required parameter & CHECK 7 (required parameters) & Pre-compilation \\
Invalid parameter value & Not caught & Runtime \\
Semantically incorrect graph & Not caught & Post-validation downstream stages \\
Unconstrained SQL string content & Not caught & Post-validation downstream stages \\
\bottomrule
\end{tabular}
\end{center}

The validator provides a hard pre-compilation guarantee for structural correctness of compiled artifacts, but not for semantic correctness of all validated plans. In the current benchmark, residual failures therefore split across two boundaries: some are caught early as structural type mismatches at validation, while others survive validation and fail later at criteria evaluation because the output contract is semantically wrong.

\subsection{Summary: Why Failures Differ}

\textbf{PlanCompiler failures are:}

\begin{itemize}
    \item Localized into two recurring boundary failures
    \item Split between late semantic-output misses and early validation-stage type mismatches
    \item Architectural in the sense that both families track specific representation boundaries
    \item Partly addressable through tighter SQL-surface constraints and clearer persistence-boundary typing
\end{itemize}

\textbf{Baseline failures are:}

\begin{itemize}
    \item Distributed (across 5+ error types)
    \item Variable (same task, different failure each run)
    \item Less directly addressable through architecture-only fixes
    \item Compound (mix of column naming, schema drift, imports)
\end{itemize}

This difference demonstrates the value of constraint enforcement: errors that are diverse and unpredictable in free-form generation become concentrated and addressable in constrained systems.

\section{Limitations}
\label{sec:limitations}

\subsection{Dependence on LLM Planning}

The system continues to rely on a large language model for initial workflow planning. While deterministic compilation improves execution reliability, incorrect plans can still appear in two observed forms: semantically wrong but executable workflows that fail only at criteria evaluation, and structurally invalid workflows rejected at validation around the SQLite persistence boundary.

Future work will explore stronger planner-side checks and bounded revision mechanisms to improve robustness.

\subsection{Absence of Automatic Repair}

The current validation layer performs static checks on graph structure, node compatibility, and parameter requirements. When validation fails, the system surfaces the error but does not yet attempt automatic plan repair.

As a result, invalid graphs terminate rather than being revised within the same run. Incorporating bounded validator-guided revision is a natural next step.

\subsection{Dependence on Node Library Coverage}

System capabilities depend on the coverage and quality of the node registry. If required functionality is not represented as an available node, the planner cannot construct a valid workflow.

The current node registry is designed primarily for data processing and data pipeline tasks. The architecture is architecturally extensible across domains, but the present registry, templates, and evaluation focus on a single workflow family.

Extending the registry to additional domains (e.g., APIs, document processing, or web automation) would better evaluate generality.

\subsection{Prompt Token Overhead}

The current implementation transmits all node descriptions to the planner, which increases prompt length and token usage as the registry grows.

Future work will explore domain-aware node selection, where the system first identifies the relevant domain and then provides only a subset of compatible nodes to the planner. This would significantly reduce token usage and improve scalability.

\subsection{Current Evaluation Scope}

The current evaluation covers ordered single-stream workflows, where nodes execute in a single directed sequence of transformations. While the graph representation allows directed edges, the current shipped validator and auto-generated execution path do not yet evaluate more complex control-flow patterns such as branching, conditional execution, parallel subgraphs, or repeated node instances.

This should be read as the scope of the present benchmark rather than as a claim about the full architectural design space. Extending the type system and execution engine to support branching graphs and richer control-flow semantics is an important direction for future work.

\section{Conclusion}
\label{sec:conclusion}

We have presented PlanCompiler, a deterministic compilation architecture for structured multi-step LLM pipelines that separates planning from execution through a typed node registry, a static validator, and a deterministic compiler. By restricting the language model, in the reported benchmark configuration, to selecting and parameterizing nodes from a fixed registry, the system enforces structural correctness before any code is emitted and substantially narrows the space of possible failures.

\subsection{Main Results}

Across a 300-task benchmark spanning shallow and deep pipelines, SQL roundtrips, and schema-themed tasks, PlanCompiler achieves substantially higher first-pass reliability than direct prompting baselines. This advantage persists as pipeline complexity increases and is especially strong on structurally demanding tasks, supporting the paperâ€™s central claim: deterministic compilation substantially reduces several classes of structural failure by construction within the current registry-constrained execution model.

\subsection{Failure Analysis and Architectural Boundary}

The most informative remaining failures are not arbitrary. They localize to two specific boundaries in the current architecture. One is a semantic-output boundary: QueryEngine accepts a semantically open SQL string, and grouped aggregation is sometimes routed through this surface rather than through the constrained Aggregator node. The other is a persistence/type boundary: after SQLite round-trips, the planner sometimes loses track of the distinction between DBHandle, file path, and materialized table. Together, these two families show that residual failures are narrow and interpretable rather than diffuse.

The evaluation also makes the validator boundary explicit. Structural failures can be caught before compilation and prevented from becoming execution-time failures, while semantically incorrect but structurally valid plans can still survive validation and fail later at criteria evaluation. The residual failures therefore do not undermine the architecture's guarantees; they locate their current limits.

Baseline models fail differently. GPT-4.1 is more prone to schema and naming errors, while Claude Sonnet 4.6 performs relatively well on SQL-heavy tasks but remains less reliable overall. These distinct failure signatures suggest that baseline reliability is not a single problem with a single remedy. PlanCompilerâ€™s contribution is not only to improve success rates, but to shift failures into a smaller, more interpretable class.

Despite added planning structure, the compiler also yields a substantially lower cost per successful task than either baseline. For structured workflow generation, deterministic compilation offers a practical path toward making LLM systems more reliable, more inspectable, and more predictable.

\subsection{Future Work}

Five directions follow naturally from this work.

First, tightening the semantically open QueryEngine surface and making the SQLite persistence boundary more explicit to the planner would likely yield the largest immediate reliability gains.

Second, extending the benchmark to additional domains would test whether the architecture transfers beyond the current data-pipeline setting. The present evaluation focuses on structured data workflows such as CSV/JSON ingestion, transformation, aggregation, and SQL persistence.

Third, planner-model ablation would test multiple language models as the planning layer to identify both the cheapest reliable planner and the most accurate one. The current implementation uses GPT-4o-mini for cost efficiency; broader comparison would make the accuracy--cost tradeoff explicit.

Fourth, adding a structured repair loop would allow the planner to revise invalid plans in response to validator feedback under bounded retry. This would address a practical limitation of the current one-shot design, although it would not remove the underlying boundary failures identified in the failure analysis.

Fifth, comparative evaluation against other structured code-generation systems, such as LangGraph, LLMCompiler, and text-to-pipeline approaches, would help locate PlanCompiler more precisely within the broader design space of constrained LLM program synthesis.

\subsection{Broader Implications}

Reliability in multi-step LLM workflows is often treated as a prompting problem---a matter of better instructions, chain-of-thought scaffolding, or output parsing. The results here suggest that, for this class of structured multi-step tasks, reliability can be productively treated as a compiler problem: identify the structural failure modes, close them by construction, and report precisely what the boundary does and does not guarantee.

The residual failures then become informative rather than opaque.

The distinction between structural and semantic correctness is the key insight. Static analysis catches structural errors by construction. Semantic errors---incorrect graph specifications, wrong parameter choices---require either human review, test-based validation, or repair loops. The architecture presented here makes this distinction explicit and measurable, enabling precise conversation about what has been gained and what remains open.

\nocite{*}
\bibliographystyle{unsrtnat}
\bibliography{references}

\appendix

\section{Full Node Registry}
\label{app:registry}

Table~\ref{tab:full_registry} provides the complete node registry with all 25 nodes across seven categories.

\begin{table}[H]
\centering
\caption{Complete node registry, 25 nodes across seven categories.}
\label{tab:full_registry}
\small
\begin{tabularx}{\textwidth}{lllll}
\toprule
\textbf{Node} & \textbf{Category} & \textbf{Input} & \textbf{Output} & \textbf{Required Params} \\
\midrule
CSVParser & Ingestion & FilePath & DataFrame & file\_path \\
JSONParser & Ingestion & FilePath & DataFrame & file\_path \\
ExcelParser & Ingestion & FilePath & DataFrame & file\_path \\
SchemaValidator & Processing & DataFrame & DataFrame & --- \\
DataTransformer & Processing & DataFrame & DataFrame & --- \\
DataFilter & Processing & DataFrame & DataFrame & condition \\
ColumnSelector & Processing & DataFrame & DataFrame & columns \\
NullHandler & Processing & DataFrame & DataFrame & strategy \\
DataSorter & Processing & DataFrame & DataFrame & by, ascending \\
TypeCaster & Processing & DataFrame & DataFrame & mapping \\
DataFrameJoin & Processing & DataFrame & DataFrame & on, how \\
StatsSummary & Processing & DataFrame & DataFrame & --- \\
DataDeduplicator & Processing & DataFrame & DataFrame & --- \\
Aggregator & Processing & DataFrame & DataFrame & group\_by, agg\_func \\
SQLiteConnector & Storage & DataFrame & DBHandle & db\_path, table\_name \\
SQLiteReader & Storage & FilePath & DBHandle & db\_path \\
PostgresConnector & Storage & DataFrame & DBHandle & connection\_string, table\_name \\
QueryEngine & Storage & DBHandle & DataFrame & query \\
CSVExporter & Export & DataFrame & FilePath & output\_path \\
JSONExporter & Export & DataFrame & FilePath & output\_path \\
RESTEndpoint & API/HTTP & DBHandle & HTTPResponse & route, port \\
AuthMiddleware & API/HTTP & HTTPResponse & HTTPResponse & api\_key\_env\_var \\
ErrorHandler & API/HTTP & ANY & HTTPResponse & --- \\
Logger & Observability & ANY & ANY & --- \\
HTTPToDataFrame & Adapter & HTTPResponse & DataFrame & --- \\
\bottomrule
\end{tabularx}
\end{table}

\textbf{Design notes.}

DataSorter requires both by and ascending as required parameters with no defaults. This was changed in v4.0 after tasks specifying descending sort silently passed \texttt{file\_has\_column} criteria while producing incorrectly ordered output. Making ascending required forces the validator to reject any plan that omits it.

SQLiteReader is an entry-point node only. Its input\_type is FilePath as a registry convention---it opens a pre-existing .db file from disk via the db\_path parameter and must only appear at the start of a pipeline. The correct write-then-query pattern is SQLiteConnector $\rightarrow$ QueryEngine. Placing SQLiteReader after SQLiteConnector produces a TYPE\_MISMATCH (DBHandle $\neq$ FilePath) which the validator correctly catches.

Aggregator always produces a column named count for count aggregations. This is a fixed property of the node implementation, not a parameter. Tasks requiring a count column must use the Aggregator node; pipelines that push GROUP BY COUNT(*) into the QueryEngine SQL string produce a column named COUNT(*) which fails \texttt{file\_has\_column: count} criteria.

Logger accepts and emits ANY, allowing it to be inserted at any point in a pipeline as a transparent observability probe without breaking type compatibility. Prior to v5.0, CHECK 3 only exempted edges where the target's input type was ANY, incorrectly raising TYPE\_MISMATCH on Logger outputs wired to typed nodes. The v5.0 fix extended the exemption to both directions.

DataFrameJoin remains in the registry as a reserved two-input primitive, but the current validator and auto-generated execution path do not support true multi-input execution. It is therefore not exercised in the reported benchmarks, which use single-stream pipelines throughout.

The API/HTTP nodes (\texttt{RESTEndpoint}, \texttt{AuthMiddleware}, \texttt{ErrorHandler}, \texttt{HTTPToDataFrame}) are included in the registry for library completeness but are not exercised in the reported benchmark suite. In the current shipped code, this surface is partly provisional: \texttt{RESTEndpoint} starts a Flask server and does not return a response object, \texttt{ErrorHandler} expects a callable, and \texttt{HTTPToDataFrame} expects an object with a \texttt{.json()} method. The benchmarked reliability claims should therefore be read primarily as applying to the ingestion, DataFrame-processing, SQLite, export, and Logger-connected pipeline surface used by the tasks.

\section{Example Compiled Artifact}
\label{app:example}

The following is a representative compiled \texttt{app.py} output for the task ``Read a CSV file, filter rows where salary exceeds 40000, store in SQLite, query the result, and export to CSV.'' The artifact is self-contained---it has no imports from the compiler or registry at runtime.

\begin{lstlisting}[language=Python, caption=Example compiled PlanCompiler artifact]
# Generated by PlanCompiler
# DO NOT EDIT MANUALLY
# --- Node: CSVParser ---
import pandas as pd
def csv_parser(file_path: str) -> pd.DataFrame:
    df = pd.read_csv(file_path)
    print(f"[CSVParser] Loaded {len(df)} rows from {file_path}")
    return df

# --- Node: DataFilter ---
import pandas as pd
def data_filter(df: pd.DataFrame, condition: str) -> pd.DataFrame:
    """
    Filters rows using pandas query syntax.
    Example condition: "age > 30 and salary < 50000"
    Node: DataFilter
    """
    filtered = df.query(condition)
    print(f"[DataFilter] Applied condition: {condition}")
    return filtered

# --- Node: SQLiteConnector ---
import sqlite3
import pandas as pd
def sqlite_connector(df: pd.DataFrame, db_path: str, table_name: str):
    """
    Stores DataFrame into SQLite and returns connection.
    Node: SQLiteConnector
    """
    conn = sqlite3.connect(db_path)
    df.to_sql(table_name, conn, if_exists="replace", index=False)
    print(f"[SQLiteConnector] Stored table '{table_name}' in {db_path}")
    return conn

# --- Node: QueryEngine ---
import pandas as pd
import sqlite3

try:
    from sqlalchemy.engine import Engine as SAEngine
    _SQLALCHEMY_AVAILABLE = True
except ImportError:
    _SQLALCHEMY_AVAILABLE = False


def query_engine(conn, query: str) -> pd.DataFrame:
    """
    Executes SQL query on a DBHandle and returns a DataFrame.
    Accepts both sqlite3.Connection (from SQLiteConnector) and
    SQLAlchemy Engine (from PostgresConnector).
    Node: QueryEngine
    """
    is_sqlite = isinstance(conn, sqlite3.Connection)
    is_sqlalchemy = _SQLALCHEMY_AVAILABLE and isinstance(conn, SAEngine)

    if not is_sqlite and not is_sqlalchemy:
        raise TypeError(
            f"[QueryEngine] Unsupported DBHandle type: {type(conn).__name__}. "
            "Expected sqlite3.Connection or SQLAlchemy Engine."
        )

    result = pd.read_sql_query(query, conn)
    print("[QueryEngine] Query executed")
    return result

# --- Node: CSVExporter ---
def csv_exporter(df: pd.DataFrame, output_path: str) -> str:
    """
    Exports DataFrame to CSV.
    Node: CSVExporter
    """
    df.to_csv(output_path, index=False)
    print(f"[CSVExporter] Exported to {output_path}")
    return output_path

# --- Execution (auto-generated) ---
if __name__ == '__main__':
    out_csv_parser = csv_parser(file_path='employees.csv')
    out_data_filter = data_filter(out_csv_parser, condition='salary > 40000')
    out_sqlite_connector = sqlite_connector(out_data_filter, db_path='out.db', table_name='employees')
    out_query_engine = query_engine(out_sqlite_connector, query='SELECT * FROM employees')
    out_csv_exporter = csv_exporter(out_query_engine, output_path='output.csv')
    print(out_csv_exporter)
\end{lstlisting}

This artifact illustrates the key properties of the execution model in the reported benchmark configuration: inter-node execution state passes through function return values, there is no shared mutable in-memory state, no global context, and no implicit passing, and the execution block is auto-generated from the validated plan with no LLM involvement.

\section{Benchmark Fixture Reference}
\label{app:fixtures}

The shipped benchmark task JSON files draw from seven generated CSV/JSON fixtures. Row counts are documented precisely because \texttt{file\_row\_count} criteria depend on them.

\begin{center}
\begin{tabular}{lll}
\toprule
\textbf{Fixture} & \textbf{Total Rows} & \textbf{Key Filter Counts} \\
\midrule
people.csv & 10 & 1 row with null salary \\
customers.csv & 20 & status = active: varies by task \\
sales.csv & 40 (38 after dedup) & revenue > 100: 27 rows \\
products.json & 25 & price > 50: 15 rows \\
employees.csv & 23 & salary > 40,000: varies by task \\
predictions.csv & 30 & score > 0.5: varies by task \\
events.json & 50 & --- \\
\bottomrule
\end{tabular}
\end{center}

In the repository checkout accompanying this paper, these CSV/JSON fixtures are produced by \texttt{benchmark/fixtures/generate\_fixtures.py} rather than committed directly. The generator also creates \texttt{existing.db} and \texttt{scores.db}, but the current benchmark task JSON files do not reference those SQLite fixtures. Tasks are parameterized against the generated CSV/JSON files at authoring time; the generator fixes the random seed and asserts the benchmark-critical row counts.

\section{Reproduction}

The complete system and benchmark are available at github.com/prnvh/plancompiler.
The experiments reported in this paper correspond to release v1.1.0 of this repository.
Release: https://github.com/prnvh/plancompiler/releases/tag/v1.1.0

To reproduce results:

\begin{lstlisting}[language=bash, caption=Reproduction steps]
git clone https://github.com/prnvh/plancompiler
cd plancompiler
git checkout v1.1.0
python -m venv .venv
source .venv/bin/activate
pip install -r requirements.txt
python benchmark/fixtures/generate_fixtures.py

cat > .env <<EOF
OPENAI_API_KEY=your_key_here
ANTHROPIC_API_KEY=your_key_here
EOF


\# Run compiler benchmark (per set)
python benchmark/harness.py --tasks benchmark/tasks/tasks\_set\_a.json --output benchmark/results/results\_set\_a.json
python benchmark/harness.py --tasks benchmark/tasks/tasks\_set\_b.json --output benchmark/results/results\_set\_b.json
python benchmark/harness.py --tasks benchmark/tasks/tasks\_set\_c.json --output benchmark/results/results\_set\_c.json
python benchmark/harness.py --tasks benchmark/tasks/tasks\_set\_d.json --output benchmark/results/results\_set\_d.json
python benchmark/harness.py --tasks benchmark/tasks/tasks\_set\_e.json --output benchmark/results/results\_set\_e.json
python benchmark/harness.py --tasks benchmark/tasks/tasks\_set\_f.json --output benchmark/results/results\_set\_f.json

\# Run baseline (per set, after compiler)

python benchmark/run\_baseline.py --tasks benchmark/tasks/tasks\_set\_a.json --results benchmark/results/results\_set\_a.json
python benchmark/run\_baseline.py --tasks benchmark/tasks/tasks\_set\_b.json --results benchmark/results/results\_set\_b.json
python benchmark/run\_baseline.py --tasks benchmark/tasks/tasks\_set\_c.json --results benchmark/results/results\_set\_c.json
python benchmark/run\_baseline.py --tasks benchmark/tasks/tasks\_set\_d.json --results benchmark/results/results\_set\_d.json
python benchmark/run\_baseline.py --tasks benchmark/tasks/tasks\_set\_e.json --results benchmark/results/results\_set\_e.json
python benchmark/run\_baseline.py --tasks benchmark/tasks/tasks\_set\_f.json --results benchmark/results/results\_set\_f.json
\end{lstlisting}

\end{document}